\documentclass[a4paper,fleqn,usenatbib]{mnras}

\usepackage{graphicx,subfig}
\usepackage[percent]{overpic}
\usepackage{lipsum}
\usepackage{multirow}
\usepackage{color}
\usepackage{rotating}
\usepackage{mwe,tikz}
\usepackage[percent]{overpic}
\usepackage{mathtools}
\usepackage{physics}
\usepackage{amsmath}
\usepackage[utf8x]{inputenc}
\usepackage{movie15}
\usepackage[T1]{fontenc}

\def\ltsima{$\; \buildrel < \over \sim \;$}
\def\simlt{\lower.5ex\hbox{\ltsima}}
\def\gtsima{$\; \buildrel > \over \sim \;$}
\def\simgt{\lower.5ex\hbox{\gtsima}}
\def\kms{{\rm\,km\,s^{-1}}}

\def\kpc{{\rm\,kpc}}


\def\deg{^\circ}

\def\etal{{\it et al.\ }}

\def\Gyr{{\rm\,Gyr}}

\def\masyr{{\rm\,mas \, yr^{-1}}}

\def\etal{{et al.~}}
\def\ltsima{$\; \buildrel < \over \sim \;$}
\def\gtsima{$\; \buildrel > \over \sim \;$}

\title[GD-1 detection using \texttt{STREAMFINDER}]{\texttt{STREAMFINDER} II: A possible fanning structure parallel to the GD-1 stream in Pan-STARRS1}

\author[Malhan \etal]{Khyati Malhan,$^{1}$\thanks{E-mail: khyati.malhan@astro.unistra.fr}
Rodrigo A. Ibata,$^{1}$
Bertrand Goldman,$^{1,2}$
Nicolas F. Martin,$^{1,2}$
\newauthor
Eugene Magnier,$^{3}$
Kenneth Chambers$^{3}$
\\
$^{1}$Universit\'e de Strasbourg, CNRS, Observatoire Astronomique de Strasbourg, UMR 7550, F-67000 Strasbourg, France\\
$^{2}$Max-Planck-Institut fur Astronomie, K{\"o}nigstuhl 17, D-69117 Heidelberg, Germany\\
$^{3}$Institute of Astronomy, University of Hawaii, 2680 Woodlawn Drive, Honolulu, Hawaii 96822, USA\\
}

\date{Accepted 2018 May 16. Received 2018 May 16; in original form 2018 April 9}

\pubyear{2017}

\begin{document}
\label{firstpage}
\pagerange{\pageref{firstpage}--\pageref{lastpage}}
\maketitle

\begin{abstract}
\texttt{STREAMFINDER} is a new algorithm that we have built to detect stellar streams in an automated and systematic way in astrophysical datasets that possess any combination of  positional and kinematic information. In Paper I, we introduced the methodology and the workings of our algorithm and showed that it is capable of detecting \textit{ultra-faint} and distant halo stream structures containing as few as $\sim 15$ members ($\Sigma_{\rm G} \sim 33.6\, {\rm mag \, arcsec^{-2}}$) in the Gaia dataset. Here, we test the method with real proper motion data from the Pan-STARRS1 survey, and by selecting targets down to $r_{0}=18.5$~mag we show that it is able to detect the GD-1 stellar stream, whereas the structure remains below a useful detection limit when using a Matched Filter technique. The radial velocity solutions provided by \texttt{STREAMFINDER} for GD-1 candidate members are found to be in good agreement with  observations. Furthermore, our algorithm detects a $\sim 40^{\deg}$ long structure approximately parallel to GD-1, and which fans out from it, possibly a sign of stream-fanning due to the triaxiality of the Galactic potential. This analysis shows the promise of this method for detecting and analysing stellar streams in the upcoming Gaia DR2 catalogue.
\end{abstract}


%%%%%%%%%%%%%%%%%%%%%%%%%%%%%%%%%%%%%%%%%%%%%%%%%%
%%%%%%%%%%%%%%%%% BODY OF PAPER %%%%%%%%%%%%%%%%%%

\section{Introduction}\label{sec:Introduction}

Stellar streams hold great promise for Galactic Archaeology \citep{Freeman2002}. Their orbital structures are sensitive tracers of galaxy formation history, the underlying Galactic potential \citep{Johnston1996,EyreBinney2009, LawMajewski2010} and the lumpiness in the dark matter distribution \citep{Ibata_2002DM_TS, Johnston2002DM_TS, StreamGap_Carlberg2012, StreamGap_Erkal2016, StreamGap_Sanders2016}. Therefore, both stream-detection and stream dynamical analysis are currently hot fields of astrophysics and small-scale cosmology.  

In the past two decades, several stellar streams have been detected around the Milky Way galaxy (\citealt{Ibata2001, Odenkirchen2001,GrillmairGD12006, Belokurov2006, GrillmairJohnson2006NGC5466, Grillmair_4_streams2009, Williams2011Aquarius, Bernard2014, Koposov_ATLAS2014, Martin_Pandas_2014, Bernard2016, Eridanus_pal15_2017,DES_Streams2017, Grillmair4New2017, CandidateGD1Pal5CPS2017,   Helmi2017_Box, Jet2017, Mateu2017, DES_11_streams2018}; see also \cite{GrillmairCarlin2016} and \cite{SmithStreamSummary2016} for recent reviews). Most streams were detected in large area surveys like SDSS \citep{SDSS2000}, Pan-STARRS1 \cite{Chambers2016PS1, PanSTARRS_Kaiser2002} and  DES \citep{DES_11_streams2018}, and a few by radial velocity surveys like RAVE \citep{Rave_Steinmetz2006} and TGAS \citep{GaiaDR12016, Lindegren2016}. The much larger number of streams found in the photometric surveys is simply a consequence of much better statistics. However, this handicap of the kinematic surveys will soon be overcome thanks to the ESA/Gaia mission \citep{Gaia2012deBruijne, GaiaDR12016,GaiaDR2_2018_Brown}, whose second data release (DR2) is scheduled for the $25^{th}$ April 2018 \footnote{Gaia DR2 shall deliver parallaxes and proper motions for all stars down to ${\rm G_0}\sim21$, and radial velocities for stars brighter than ${\rm G_0}\sim13$}. It is likely that the Milky Way contains a large number of hitherto undetected stellar streams. With the exceptional quality of the data expected in Gaia DR2, it is clearly worthwhile to devote effort to design new and better stream detection schemes.

Existing stream detection techniques employ data analysis methodologies that exploit only a subset of the information that we will soon have on large numbers of stars in the Milky Way. For example, the Pole Count technique -\citep{Johnston1996, Ibata2002PoleCount} utilizes only the positions of the stars, whereas the Matched-Filter technique \citep{Balbinot2011MF, Rockosi2002}) employs only the positions and the photometry of the stars. Most of the  detections of co-moving groups have been made on the basis of structural coherence by looking for clumping of stars only in  velocity space \citep{Helmi2017_Box, Virgo_Duffau_2006, Aquarius_Williams_2011ApJ...728..102W}. Stream detection techniques engaging all the stellar information (positions, kinematics and photometry) simultaneously should definitely improve the scope and the significance of stream detection. It has been long suggested that the right set of coordinates to identify substructures is the space of the integrals of motion, but that requires complete knowledge of the 6D phase-space distribution of the stars, something that even Gaia will only deliver for bright and nearby objects (a very small subset of the full Gaia sample, see \citealt{CFIS_I_2017}). Therefore, stream search methods using integrals of motion will not be very useful for detecting streams that exist in the distant halo of the Milky Way.

Given the quality of the data that shall soon become available from Gaia DR2, and what we perceived as the shortcomings of the existing stream detection techniques, we decided to build a new stream detecting algorithm (the \texttt{STREAMFINDER}; \citealt{Malhan2018_SF}, hereafter Paper~I). The main purpose of \texttt{STREAMFINDER} is to detect cold and narrow tidal stellar streams, of the type produced by the tidal disruption of globular clusters or very low-mass galaxies. The algorithm incorporates our prior knowledge of stellar streams and analyses data in position, kinematics and photometry space simultaneously to maximize the stream detection efficiency. Our algorithm makes use of the realization that the members of a stream can be contained within a 6D hyper-dimensional tube (or hypertube) that coils much like an orbit in phase-space, with width in real and velocity space similar to the size and velocity dispersion of the progenitor cluster. For each star (with acceptable proper motion measurements), the algorithm shoots orbits using the phase-space information of the data, as observed today, in a realistic Milky Way potential. These orbits are transformed into hypertubes within the algorithm, with plausible phase-space width and length. The number of stars that get encapsulated within these hypertubes are then counted. After processing all the survey stars in this manner, the output of the algorithm can be summarized in a density plot where the likelihood of every star corresponds to how well a star is coherently compatible with other stars in the data to form a stream like structure (see Figure~7 of Paper~I). 

In Paper~I, we introduced the \texttt{STREAMFINDER} algorithm that we have built, explained the physical motivation behind it and demonstrated its workings. Our analysis, based on a mock dataset of the quality Gaia will deliver suggested that the algorithm is capable of detecting even ultra-faint stream features lying well below previous detection limits. Our tests showed that the algorithm will be able to detect distant halo stream structures $>10^{\circ}$ long containing as few as $\sim 15$ members ($\Sigma_{\rm G} \sim 33.6\, {\rm mag \, arcsec^{-2}}$) in the Gaia dataset. 

Motivated by these results, and to test the reliability of the machinery that we have built on a real dataset, we apply it here to the Pan-STARRS1 survey, in a region containing the so called ``GD-1'' stream. The GD-1 stream is a quintessential example of a dynamically cold and narrow stream structure extending over $60^{\circ}$ on the sky and devoid of any significant internal velocity dispersion or distortion \citep{GrillmairGD12006, CarlbergGD1paramter2013}. It is a perfect example of the class of stream structure \texttt{STREAMFINDER} is constructed to find. 

This paper is arranged as follows. In Section \ref{sec:Data_Analysis} we briefly describe the data used. Section \ref{sec:GD-1_stream_detection} shows the detection of the GD-1 stream using the Matched Filter technique. In Section \ref{sec:Comparison} we make the comparison between our algorithm \texttt{STREAMFINDER} and the Matched Filter. In Section \ref{sec:new_stream} we discuss the discovery of a new structure that we have found in the neighbourhood of GD-1. In Section \ref{sec:Radial_velocity} we present additional power that our algorithm holds in predicting the missing phase-space information of the detected stream structures. Finally, in Section \ref{sec:Discussion_and_Conclusions} we discuss our results.

% %%%%%%%%%%%%%%%%%%%%%%%%%%%%%%%%%%%%%%%%%%%%%%%%%%%%%%%%%%%%%%%%%%%%%%%%%%%%%%%
% % Section
% %%%%%%%%%%%%%%%%%%%%%%%%%%%%%%%%%%%%%%%%%%%%%%%%%%%%%%%%%%%%%%%%%%%%%%%%%%%%%%%
%https://arxiv.org/pdf/1703.06278.pdf. See page 11 of thhis paper to know what red and black line I am talking about.
\begin{figure*}
\begin{center}
\includegraphics[width=\hsize]{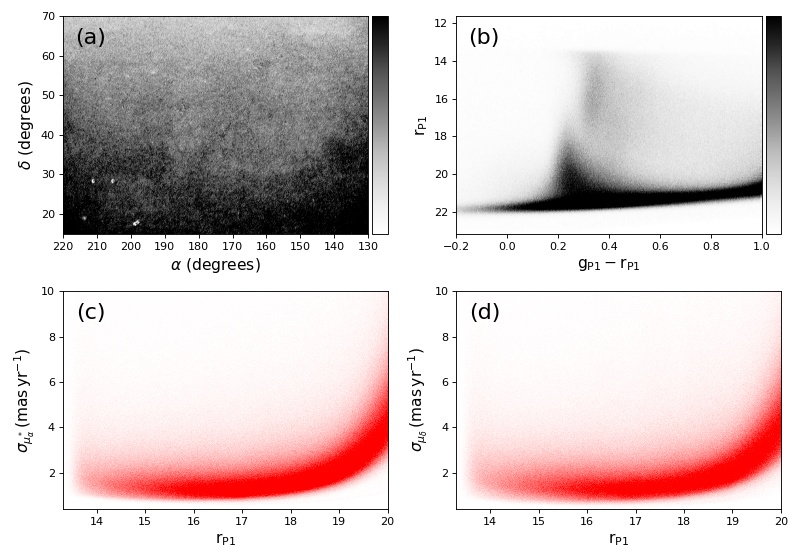}
\end{center}
\caption{PS1 proper motion dataset. (a) Raw density map of stars in the given patch of the sky obtained using the PS1 catalogue. The darker regions imply higher density regions. The GD-1 stream lies in this particular area of the sky. (b) Colour-magnitude Hess-Diagram of the same patch of sky. The stars on the red side of ${\rm g_{P1}-r_{P1}}$ = 1.0 consists mostly of local M dwarfs of the disk and are not used in our analysis. Panels (c) and (d) show the behaviour of the proper motion uncertainties with respect to ${\rm r_{P1}}$. These uncertainties become very large for ${\rm r_{P1}} > 20$.}
\label{fig:global_PS_data}
\end{figure*}

\section{Data Analysis} \label{sec:Data_Analysis}

We use the Pan-STARRS1 (PS1) proper motion dataset \citep{PanSTARRS_Kaiser2002,  Kaiser2010PS1, Tonry2012PS1, Chambers2016PS1, Magnier2016PS1} in all of the present analysis. In terms of astrometry, PS1 delivers 2D positions and 2D proper motions of the stars along with photometry in five bands ( ${\rm g_{P1} ,r_{P1} ,i_{P1}, z_{P1}, y_{P1}}$) %BG I put the filter in math mode, see if you like it
with a 5$\sigma$ single epoch depth of (23.3, 23.2, 23.1, 22.3, 21.4). At present, the PS1 survey is the only dataset that delivers proper motions for all stars with $\delta > -30^{\circ}$ down to ${\rm r_{P1}}\sim23.0$. In terms of kinematic quality, this dataset will be soon be superseded by Gaia DR2 (for those stars with ${\rm G_o} < 21$), nevertheless, the dataset provides us a unique opportunity to test the functionality and feasibility of our algorithm on an actual astrophysical catalogue before Gaia DR2 becomes available. But even after the Gaia release, this dataset will still remain highly useful for analysing stars fainter than the Gaia detection limit.

Each image of PS1 is calibrated against the Gaia DR1 position. To this end, the Gaia position of the astrometric reference stars are propagated back to the PS1 image epoch, using a  model describing the Galactic rotation and the Solar motion and using the photometric distance to the reference stars; see \cite{Green2015DustDerredening} and \citet{Magnier2016PS1}, respectively. If that model were a perfect description of the motions of the Galaxy and the Sun, the resulting PS1 proper motions and parallaxes obtained by fitting the PS1, 2MASS and Gaia (if available) positions would be inertial and extragalactic objects would have null proper motions and parallaxes. To confirm this, we first selected a sample of galaxies using both the light profiles of the objects, and their colours. Specifically, we required that the difference $m_{\rm PSF}-m_{\rm aperture}$ be larger than 0.2\,mag {\em on average} for the four filters ${\rm g, r, i}$ and ${\rm z}$ and be all smaller than 0.5\,mag, with a signal-to-noise ratio larger than 20. In addition, we required an infrared colour of ${\rm J-W1}>1.7$\,mag where W1 is the WISE 3.4-$\mu$m magnitude, limited to the 12.0 to 15.2-mag range, and ${\rm J}$ is the UKIDSS 2-arcsec-radius aperture magnitude if available ($SNR>7$), or the 2MASS $J$ magnitude otherwise ($SNR>4$) \citep{Kovacs2015}. Finally, for each equatorial hemisphere, we calculated on a 1024x1024 grid the 3-$\sigma$-clipped mean (iterated 3 times) of the galaxies' proper motions over a box of $6\deg$ on a side.

Over the area of interest to this study, about 90~galaxies were used for each grid point. The corrections for $\mu_\alpha$ and $\mu_\delta$ vary smoothly from $-0.5$ to $+2.4$\,mas/yr and from $+2.0$ to $+3.4$\,mas/yr respectively. For each catalogue entry, we subtracted the values taken at the nearest grid point from the measured proper motions to obtain inertial proper motions.

We select the rectangular patch of PS1 sky with $130^{\circ} < \alpha < 220^{\circ}$ and $15^{\circ} < \delta < 70^{\circ}$, which corresponds to the region where GD-1 stream lies. The stars in this selected part of the sky were corrected for foreground reddening using the 3D extinction map provided by \cite{Green2015DustDerredening} \footnote{{\tt http://argonaut.skymaps.info/}}. The de-reddended and proper motion corrected data is shown in Figure \ref{fig:global_PS_data}.

% %%%%%%%%%%%%%%%%%%%%%%%%%%%%%%%%%%%%%%%%%%%%%%%%%%%%%%%%%%%%%%%%%%%%%%%%%%%%%%%
% %\section
% %%%%%%%%%%%%%%%%%%%%%%%%%%%%%%%%%%%%%%%%%%%%%%%%%%%%%%%%%%%%%%%%%%%%%%%%%%%%%%%

\begin{figure*}
\begin{center}
\vbox{
\hbox{
\includegraphics[width=\hsize]{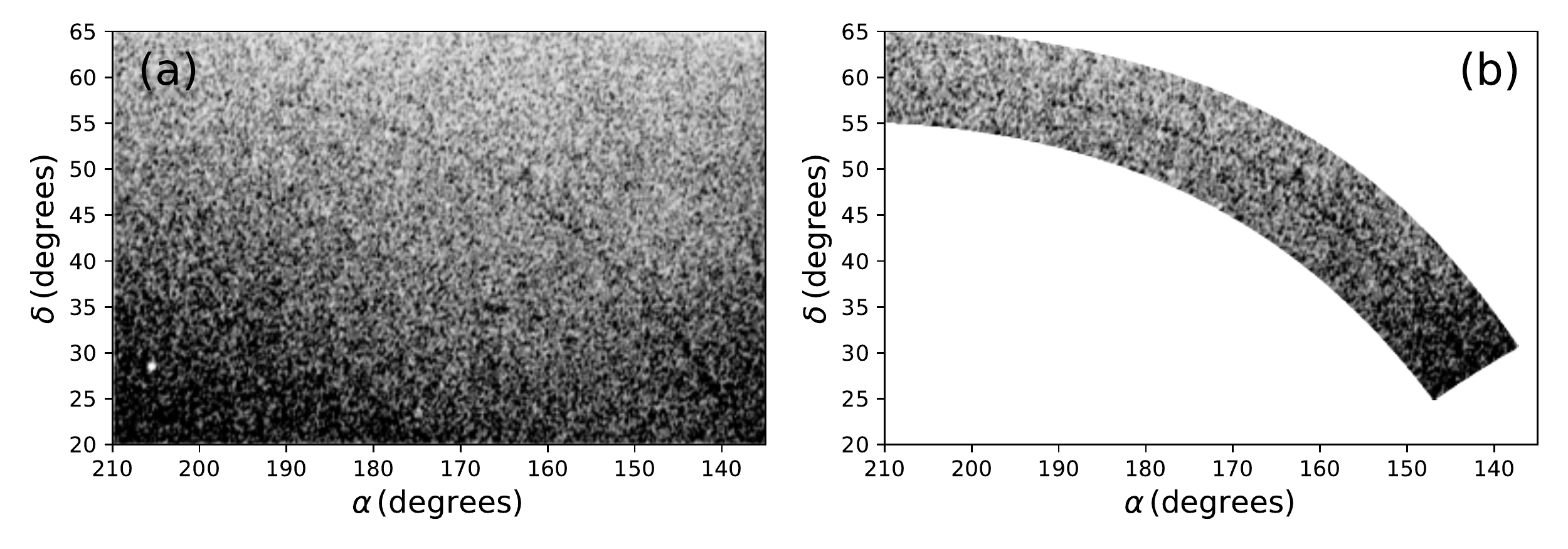}
}
}
\vbox{
\hbox{
\vspace{2.0cm}
\includegraphics[width=\hsize]{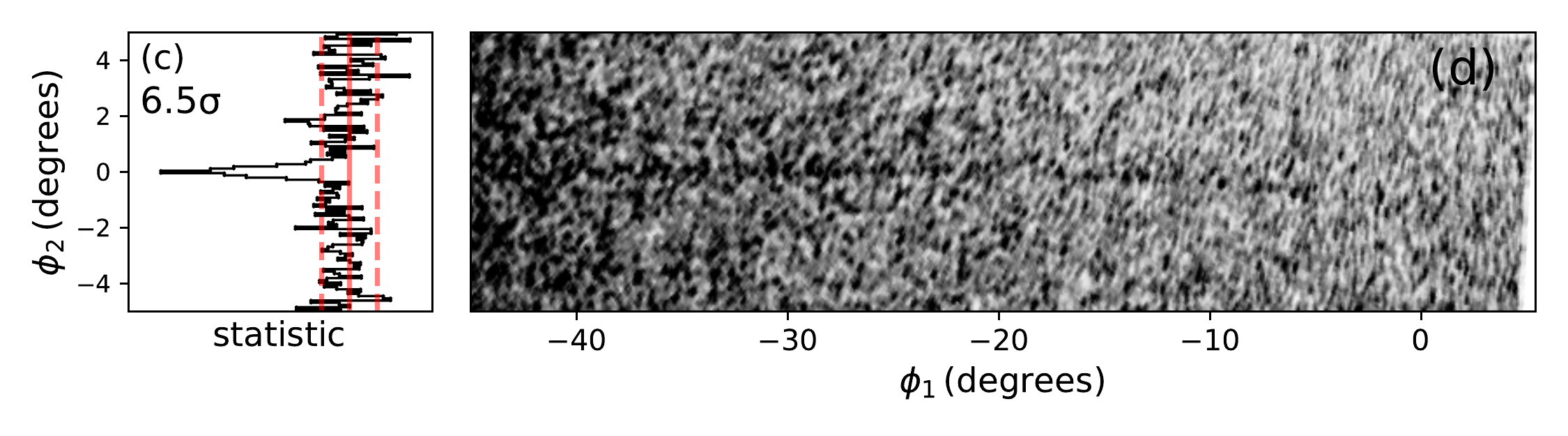}
}
}
\end{center}
\vspace{-2.50cm}
\caption{GD-1 stream detection using the MF technique. (a) MF density map of the chosen patch of sky in PS1, derived using the M13 globular cluster as the CMD template. The GD-1 stream can be seen as a $\sim 60^{\circ}$ extended structure on the sky. All stars with $14 < {\rm r_{P1}} < 21.5$ and $0.1 < {\rm g_{P1}-r_{P1}} < 0.6$ were used to create this density plot. (b) The same MF density plot is now shown in a particular area of the sky that runs along the GD-1 stream. (d) The MF plot is represented in the rotated spherical coordinate system aligned approximately with the GD-1 stream. The GD-1 stream can be seen to lie along $\phi_2 \sim 0$ in this plot. (c) We estimate that the stream is detected at a significance level of $\approx 6.5\sigma$.}
\label{fig:Our_MF_Grillmair_like}
\end{figure*}

\section{Detection of GD-1 using a Matched Filter}\label{sec:GD-1_stream_detection}

We first present the detection of the GD-1 stream using the matched filter (MF) technique, the method originally used for its detection \citep{GrillmairGD12006}. MF \citep{Rockosi2002, Balbinot2011MF} is an optimal contrast adjusting technique that relies on the colour-magnitude information of the stars. The technique works by selecting a suitable colour-magnitude diagram (CMD) single stellar population (SSP) template model that represents the stellar stream members to be detected. For many halo streams the discriminating power of the MF resides mainly at the main-sequence turnoff (MSTO) and below where the stellar density rapidly increases and where it also lies blueward of the contaminating foreground population. The GD-1 stream was initially discovered in the density plot that was obtained as a result of  the MF prepared using the CMD of the M13 globular cluster. 

To reproduce the GD-1 detection with the MF method, we first impose %BG I think it is usual to use the preteritum
CMD cuts to retain the upper main-sequence region $0.1 \le {\rm g_{P1}-r_{P1}} \le 0.6$, and trim the data below ${\rm r_{P1}}\sim21.5$. We call the resulting subset Dataset 1.

We created a MF following the procedure described in \citet{Balbinot2011MF}, and using the CMD of M13 globular cluster as the target template (similarly to \citealt{GrillmairGD12006}). We divided the CMD into bins of $0.01$~mag in colour, and $0.1$ in magnitude as well as $0.1^{\circ}$ spatially on the sky. The resulting weighted image was then smoothed with a Gaussian kernel with $\sigma=0.2^{\circ}$. 

The spatial density map thus obtained is shown in Figure~\ref{fig:Our_MF_Grillmair_like}a. Since it is convenient to work in the spherical coordinate which is aligned with the GD-1 stream, we made use of the rotation matrix provided by \cite{Koposov2010} to make a transformation of coordinates from equatorial to these new spherical coordinates. A similar MF density plot is also shown in this new rotated spherical system in Figure \ref{fig:Our_MF_Grillmair_like}d. Note that the GD-1 stream is  visible as a high contrast stream feature at a detection level of $\approx 6.5\sigma$.

% %%%%%%%%%%%%%%%%%%%%%%%%%%%%%%%%%%%%%%%%%%%%%%%%%%%%%%%%%%%%%%%%%%%%%%%%%
% % SECTION
% %%%%%%%%%%%%%%%%%%%%%%%%%%%%%%%%%%%%%%%%%%%%%%%%%%%%%%%%%%%%%%%%%%%%%%%%%
\section{Comparison between \texttt{STREAMFINDER} and Matched Filter }\label{sec:Comparison}

We deem it most useful to compare the \texttt{STREAMFINDER} to the MF technique. This is because, first of all, the GD-1 stream is too faint to be detected by a simple pole count. Secondly, the proper motion uncertainties in the PS1 dataset are too large for the GD-1 stream to be detected by using analyses that only incorporate the stellar velocity information. Moreover, most of the known Milky Way streams, like the Pal-5 stream \citep{Odenkirchen2001}, the NGC 5466 structure \citep{GrillmairJohnson2006NGC5466}, the Orphan stream \citep{Grillmair2006Orphan, Belokurov2006}, Lethe, Cocytos, and Styx \citep{Grillmair_4_streams2009}, Indus, Ravi, Jhelum, Chenab \citep{Shipp2018} and others, along with GD-1, were detected via an application of the MF technique, which demonstrates its power for stream detection.

The MF technique is expected to fail in detecting streams broadly in two cases, (1) if the stream happens to be elongated along the line of sight, or (2) if the stream is too low in contrast. The first case depends on the nature of the stream, however the second case is what can be examined here to compare the MF to \texttt{STREAMFINDER}. A stream could be observed to be low in contrast because (1) it is an ancient structure that is now very spread-out spatially, or (2) it is distant in the halo and hence its MSTO, where the majority of stream stars are expected to lie, lies below the photometric limit of the survey. Indeed, there could be many faint Milky Way stellar streams that exist in the halo but which have remained undetected by MF-based weighting techniques due to the above-mentioned reasons. \texttt{STREAMFINDER} combats this low density problem by performing a multidimensional analysis of the stars, incorporating all the stellar information in terms of positions, kinematics and photometry that in turn improves the stream detection efficiency. 

\begin{figure*}
\begin{center}
\vbox{
\includegraphics[width=\hsize]{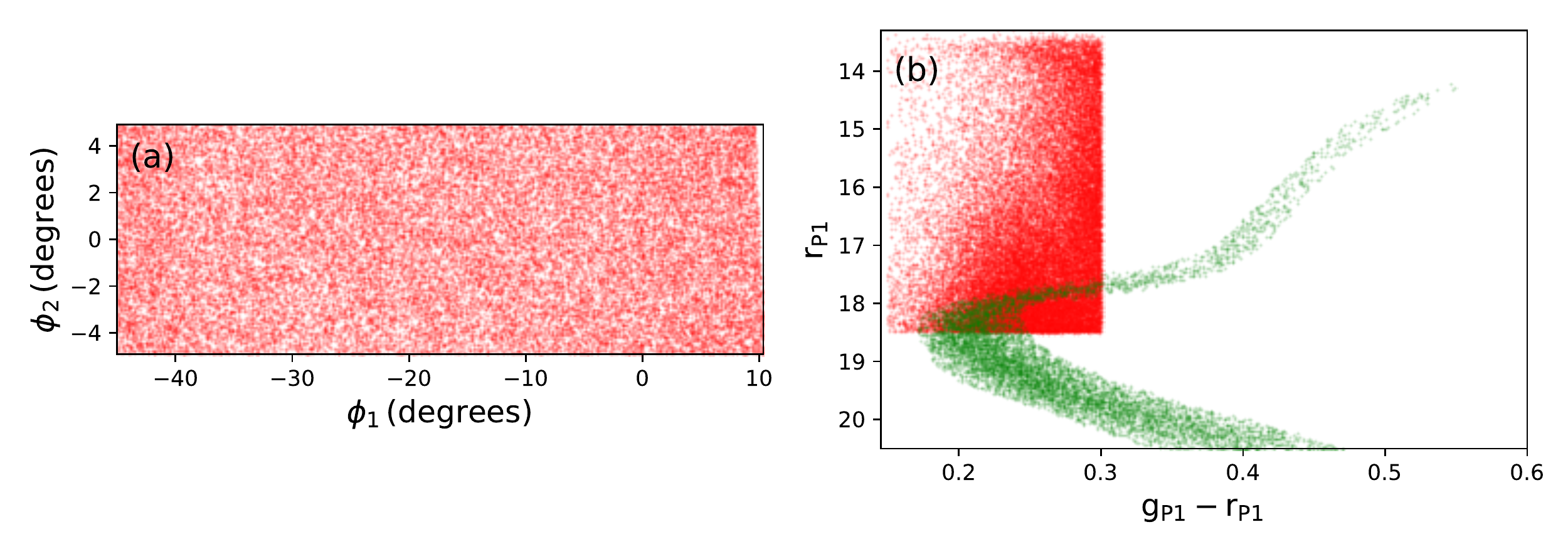}
}
\end{center}
\caption{Dataset 2. To make a comparison between \texttt{STREAMFINDER} and the MF technique we take a subset of the PS1 data around GD-1, selecting stars with ${\rm r_{P1} < 18.5}$ and $0.15 < {\rm g_{P1}-r_{P1}} < 0.30$. (a) The  subset is presented in the rotated coordinate frame. GD-1 lies at $\phi_2\sim 0$ in this frame. (b) Represents the CMD of this data subset. The green dots correspond to the M13 globular cluster CMD, originally used to detect the GD-1 stream. The plot shows that most of the MSTO stars in the GD-1 are lost, due to the colour-magnitude selection window.}
\label{fig:base_data_2}
\end{figure*}

While we cannot alter the physical structure of GD-1, we can legitimately simulate making it harder to detect by artificially reducing the limiting magnitude of the survey. 
To this end, we select only those stars in Dataset 1 that are brighter than ${\rm r_{P1} = 18.5}$ and follow the criterion $0.15 < {\rm g_{P1}-r_{P1}} < 0.30$. The colour-cut follows the selection made by \citet{Koposov2010}. We refer to this truncated sample as Dataset 2, which is shown in Figure \ref{fig:base_data_2}. While a redder colour cut would have included some GD-1 sub-giants, it would also have given rise to a greater contamination fraction, lowering the significance of the detection.

\begin{figure*}
\begin{center}
\vbox{
\includegraphics[width=\hsize]{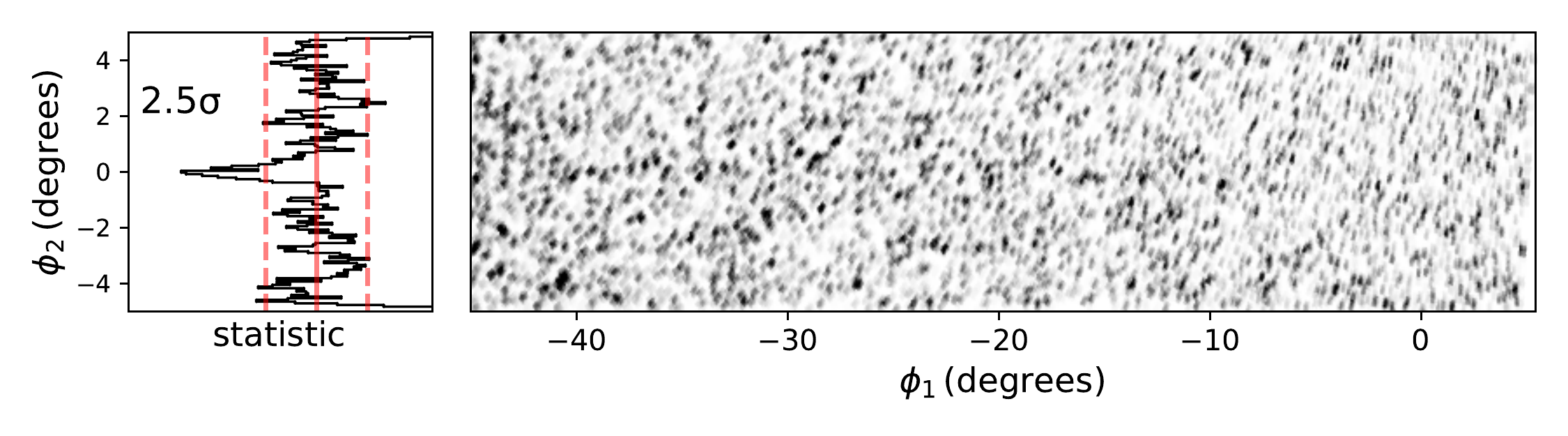}
}
\end{center}
\caption{Non-detection of GD-1 in Dataset 2 using the MF technique. The GD-1 stream now appears with only $\approx 2.5\sigma$ detection significance, due to the absence of stars fainter than ${\rm r_{P1} = 18.5}$.}
\label{fig:No_MF_detection}
\end{figure*}

\subsection{Matched Filter -- once again}

We execute the MF technique once again using Dataset 2. The resulting density plot is shown in Figure \ref{fig:No_MF_detection}. GD-1 does not appear with a very strong detection significance, a nearly non-detection with a significance of $\approx 2.5\sigma$. This is not surprising as most of the stars which lie at the MSTO and below in GD-1 were discarded while constructing the Dataset 2. The stars that received higher weights based on the MF weighting scheme are now less in number and the contrast of the GD-1 structure in the density plot is much diminished.

\subsection{Positive detection of GD-1 in Dataset 2 with \texttt{STREAMFINDER}}

We now feed the very same Dataset 2 to \texttt{STREAMFINDER}. The algorithm uses the positions and proper motions of the stars to sample orbits. For the purpose of integrating these orbits, we use only those stars for which
\begin{equation}
\sigma_{\mu_{\alpha}} < 3.0 \masyr \, \textrm{and} \, \sigma_{\mu_{\delta}} < 3.0 \masyr \, ,
% \sigma_{\mu} =\sqrt{\sigma^2_{\mu_{\alpha}} + \sigma^2_{\mu_{\delta}}} < 3.0 \masyr \, ,
\end{equation}
so that the obtained orbital solutions can be trusted. A proper motion of $\sigma_{\mu}=3\masyr$ at a distance of $10\kpc$ corresponds to an uncertainty in the transverse velocity of $\sim 150\kms$, which is already a huge uncertainty compared to expected relative motion of the stream and the contaminating population. However, once the hypertubes are calculated, we use the full Dataset 2 sample to count the number of stars that lie within the hypertubes, and to calculate the corresponding likelihood values.

The orbits are integrated within the Galactic potential model 1 of \cite{Dehnen1998Massmodel}, and these orbits are then projected into the heliocentric frame of observables. For this, we assume a Galactocentric distance of the Sun of $8.5\kpc$ and adopt the peculiar velocity of the Sun $\bmath{V_{\odot}} = (u_{\odot}, v_{\odot}, w_{\odot}) = (11.1, 12.24, 7.25) \kms$ \citep{Schornich2010_Sun}. Moreover, our algorithm uses a pre-selected isochrone model in order to sample orbits in distance space, as explained in Paper~I. The selected isochrone model essentially corresponds to the proposed SSP of the stream. For this, we choose an isochrone  with metallicity ${\rm [Fe/H]=-1.4}$ and age $9\Gyr$ \citep{Koposov2010} from the Padova stellar population models \citep{Marigo2008Padova}. This isochrone model matches  well the CMD of M13 cluster and hence that of GD-1. Other parameter ranges used to integrate orbits in the Galaxy were identical to those detailed in Paper I.

\begin{figure*}
\begin{center}
\vbox{
\includegraphics[width=\hsize]{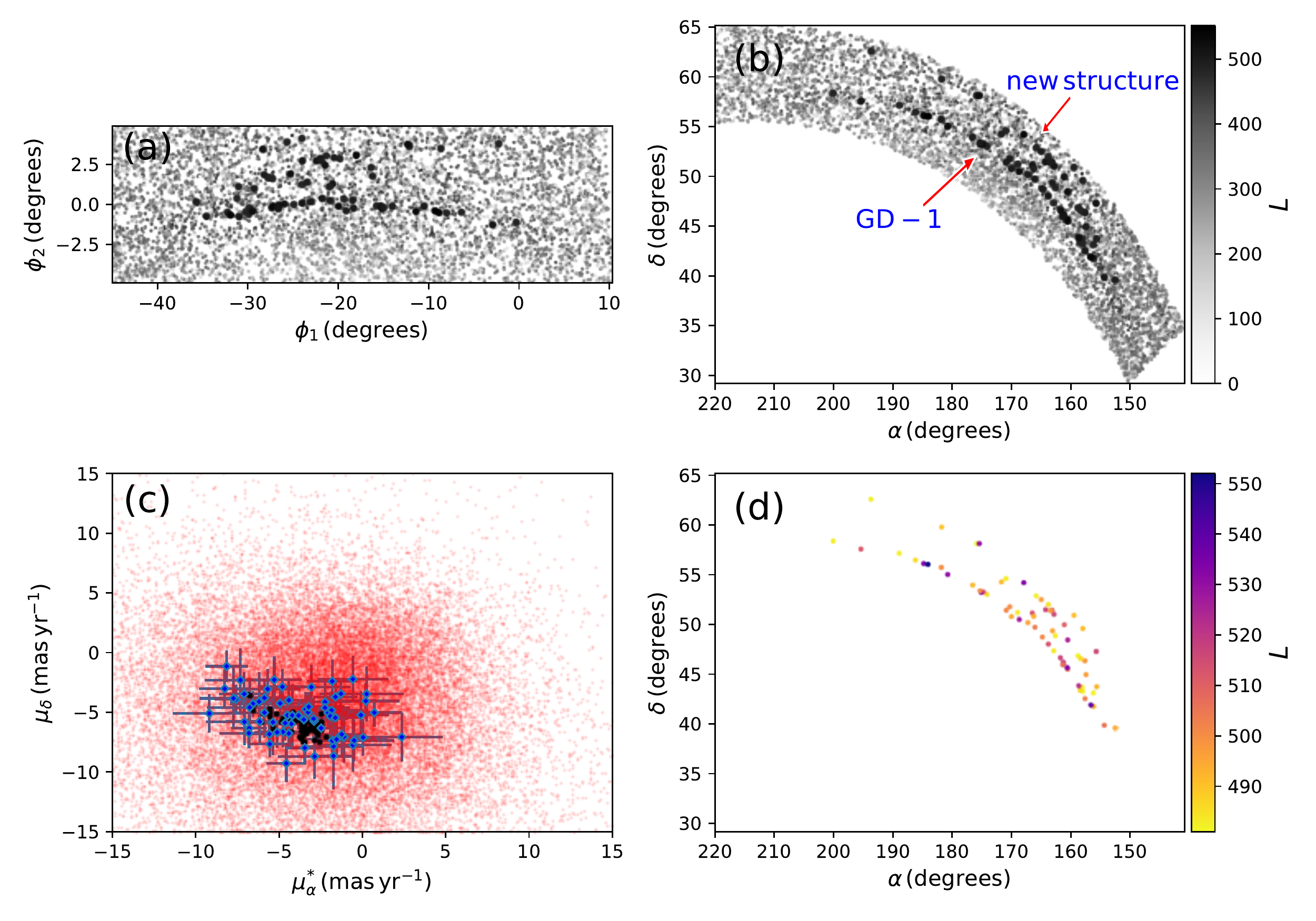}
}
\end{center}
\vspace{-0.5cm}
\caption{Detection of GD-1 with the \texttt{STREAMFINDER} in Dataset 2. (a) The algorithm after processing the given patch of sky returns a density plot that is shown here in the rotated coordinate system. The highest likelihood stars (0.2\% of the sample) are marked in large black dots that immediately reveal the GD-1 structure along $\phi_2\sim$0. Along with GD-1, \texttt{STREAMFINDER} reveals another stream feature towards the north. (b) The same as (a) but in equatorial coordinates. (c) The proper motions of the stars of Dataset 2 are shown in red. The highest likelihood stars are marked with blue dots together with their error bars. The black coloured dots show the expected proper motion values of these highest likelihood stars, a by-product of \texttt{STREAMFINDER}. (d) The highest likelihood stars are shown in equatorial coordinates, revealing the two distinct structures that are detected at comparable significance. Based on the statistics of the contamination, \texttt{STREAMFINDER} detects both stream-like structures at a $>4.4\sigma$ level of confidence.}
\label{fig:STREAMFINDER}
\end{figure*}

The spatial distribution of stream likelihood calculated by the \texttt{STREAMFINDER} is shown in Figure \ref{fig:STREAMFINDER}. Unlike the MF result from the same sample (Dataset 2), one can now vividly see the GD-1 structure, which is detected at the $\approx 4.4\sigma$ level of confidence. This means that the multidimensional analysis done by \texttt{STREAMFINDER} easily allows it to detect stream structures that would otherwise be lost with a MF search. This detection of an extremely low contrast stream shows the power of our algorithm over the MF and hence over many other stream-detection techniques. Moreover, our algorithm makes sure that the detected structure is in fact stream-like --- spatially extended and coherent in velocity space (as can be seen in Figure \ref{fig:STREAMFINDER}). 

The candidate members of the GD-1 stream identified by the \texttt{STREAMFINDER} are listed in Table \ref{tab:GD-1_members}.

\begin{table}
\caption{Highest likelihood stars candidates in PS1 along the GD-1 track, as obtained by the \texttt{STREAMFINDER}. Columns 1 and 2 list the sky positions and columns 3 and 4 are the $\rm{g_{P1}}$and $\rm{r_{P1}}$ magnitudes (the median of the PS1 measurements).}
\label{tab:GD-1_members}
\begin{center}
\begin{tabular}{cccc}
\hline
$\rm{RA}$ & $\rm{DEC}$ & $\rm{g_{P1}}$ & $\rm{r_{P1}}$ \\
 (deg) & (deg) & & \\
\hline
 154.33723& 39.82138&  18.44& 18.18 \\
 161.73164& 46.59564&  18.40& 18.19 \\
 157.58469& 42.49022&  18.47& 18.21\\ 
 158.63541& 43.79097&  18.59& 18.30 \\
 160.50852& 45.44162&  18.50& 18.25 \\
 160.56969& 45.60858&  18.54& 18.25 \\
 161.22963& 46.14118&  18.38& 18.13 \\
 161.25371& 45.96201&  18.54& 18.27 \\
 161.30762& 45.89013&  18.16& 17.87  \\
 156.59654& 41.85462&  18.49& 18.24 \\
 163.75657& 47.96208&  18.31& 18.04  \\
 165.99967& 49.65492&  18.32& 18.02 \\
 168.64583& 50.43288&  18.32& 18.08  \\
 174.67653& 53.19757&  18.20& 17.91 \\
 175.04661& 53.14976&  18.45& 18.20  \\
 184.73228& 56.03422&  18.37& 18.11 \\
 184.00126& 55.96127&  18.42& 18.19 \\
 180.68427& 54.94157&  18.17& 17.91 \\
 195.29246& 57.48671&  18.38& 18.09 \\
\hline
\end{tabular}
\end{center}
\end{table}

\begin{table}
\caption{As Table~\ref{tab:GD-1_members}, but for the highest likelihood stars obtained by the \texttt{STREAMFINDER} selected along the structure that appears parallel to GD-1.}
\label{tab:Hanawi_members}
\begin{center}
\begin{tabular}{cccc}
\hline
$\rm{RA}$ & $\rm{DEC}$ & $\rm{g_{P1}}$ & $\rm{r_{P1}}$ \\
 (deg) & (deg) & & \\
\hline
 155.69694& 47.22650&  18.54& 18.31\\
 160.48647& 48.38166&  18.61& 18.38\\
 161.06528& 49.90054&  18.36& 18.07\\
 167.88944& 54.13320&  18.44& 18.19\\
 162.80495& 50.95156&  18.39& 18.08\\ 
 164.22409& 51.40948&  18.53& 18.29\\
 163.13786& 51.35826&  18.45& 18.17\\
 175.35981& 58.04287&  18.29& 18.02\\
\hline
  
\end{tabular}
\end{center}
\end{table}

% %%%%%%%%%%%%%%%%%%%%%%%%%%%%%%%%%%%%%%%%%%%%%
% % Section
% %%%%%%%%%%%%%%%%%%%%%%%%%%%%%%%%%%%%%%%%%%%%%
\section{Probable fanning of the GD-1 stream}\label{sec:new_stream}

The likelihood distribution plot shown in Figure \ref{fig:STREAMFINDER} reveals another stream-like feature alongside GD-1. We find that the significance of this structure is comparable to that of GD-1, appearing at a detection level of $> 4.4\sigma$. \citet{GrillmairGD12006} mention in passing that ``There may be a second, more diffuse feature with $174\deg<\alpha<200\deg$ about $3\deg$ to the north of [GD-1]''. Here we confirm the detection of the feature at a level of significance sufficient to confirm its discovery. 
The structure appears to be extended over a length of $\sim 40^{\deg}$ from $155^{\deg}<\alpha<195^{\deg}$. 

\begin{figure*}
\begin{center}
\vbox{
\includegraphics[width=\hsize]{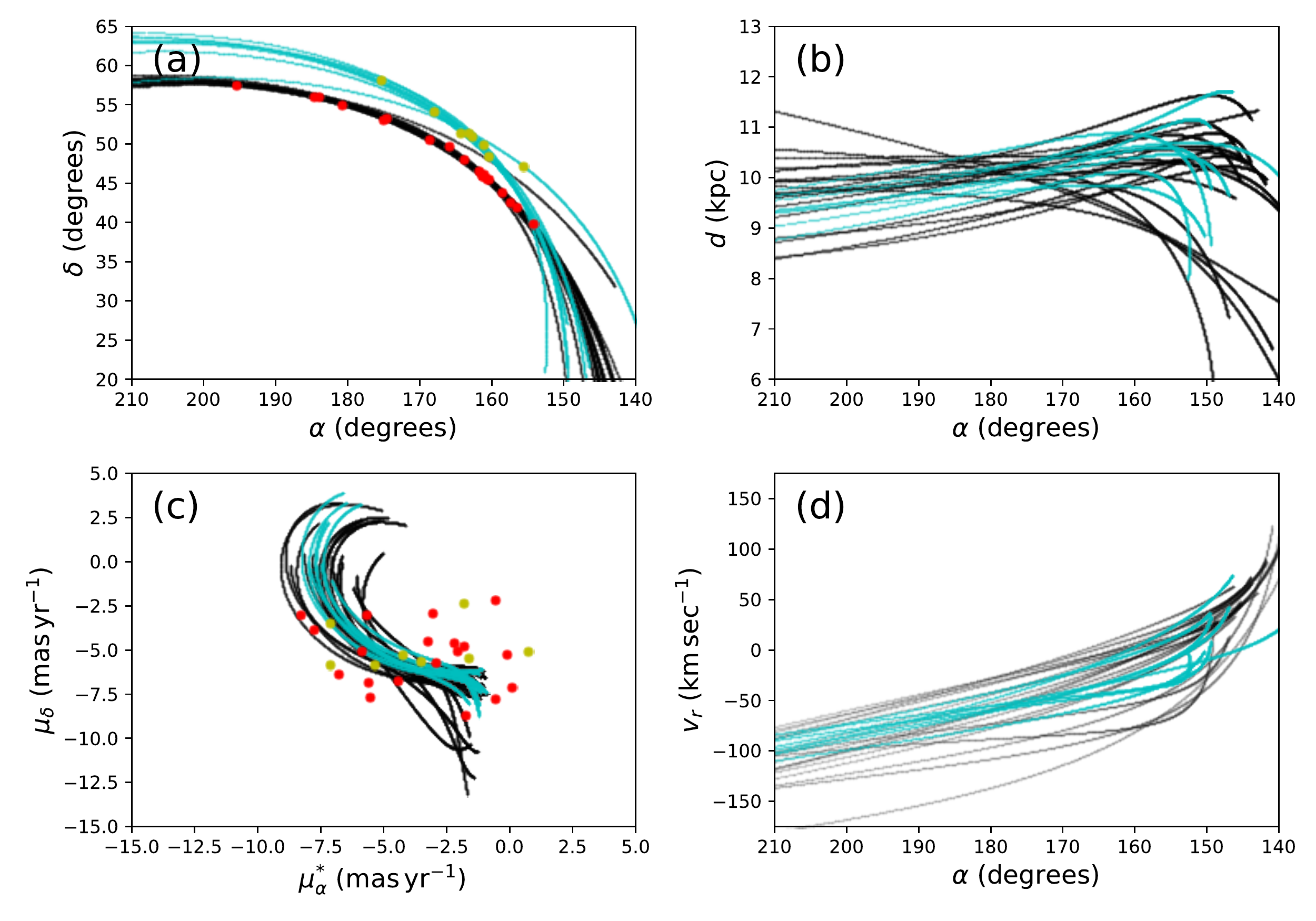}
}
\end{center}
\vspace{-0.5cm}
\caption{Orbital solutions of GD-1 and the parallel structure. The 27  stars with highest likelihood are plotted here in dots red (for GD-1) and yellow (for the feature to the north). We use their orbital solutions, obtained as a by-product from \texttt{STREAMFINDER}, to make a comparison with the observations. The top left (bottom left) plot compares the orbits with these data points in position (proper motion) space. The orbits obtained from GD-1 and the parallel structure candidate members are shown in black and cyan colours, respectively. The top right (bottom right) plot shows the behaviour of the orbits of the two streams in distance (radial velocity) space. Note the overlapping of the orbits at $(\alpha, \delta)\sim(152^{\deg},38^{\deg})$.}
\label{fig:orbital_solutions}
\end{figure*}

In Figure \ref{fig:orbital_solutions}, we show possible orbital solutions for both GD-1 and this additional structure that we obtain as a by-product of \texttt{STREAMFINDER} (see Paper~I). Although the two features appear as clearly-distinguishable stream-like structures on the sky (Figure \ref{fig:orbital_solutions}a), interestingly, their orbits seem to overlap in distance and velocity space.

At the distance of GD-1, the PS1 proper motion uncertainties correspond to a typical uncertainty on the transverse motion of $>50\kms$. This, together with the absence of radial velocity measurements, makes it hard to speculate on the orbital properties of the system at this stage. The possible candidate members of the structure parallel to GD-1 are listed in Table \ref{tab:Hanawi_members}.

%%%%%%%%%%%%%%%%%%%%%%%%%%%%%%%%%%%%%%%%%%%%%%
% Section
%%%%%%%%%%%%%%%%%%%%%%%%%%%%%%%%%%%%%%%%%%%%%%
\begin{figure}
\begin{center}
\vbox{
\includegraphics[width=\hsize]{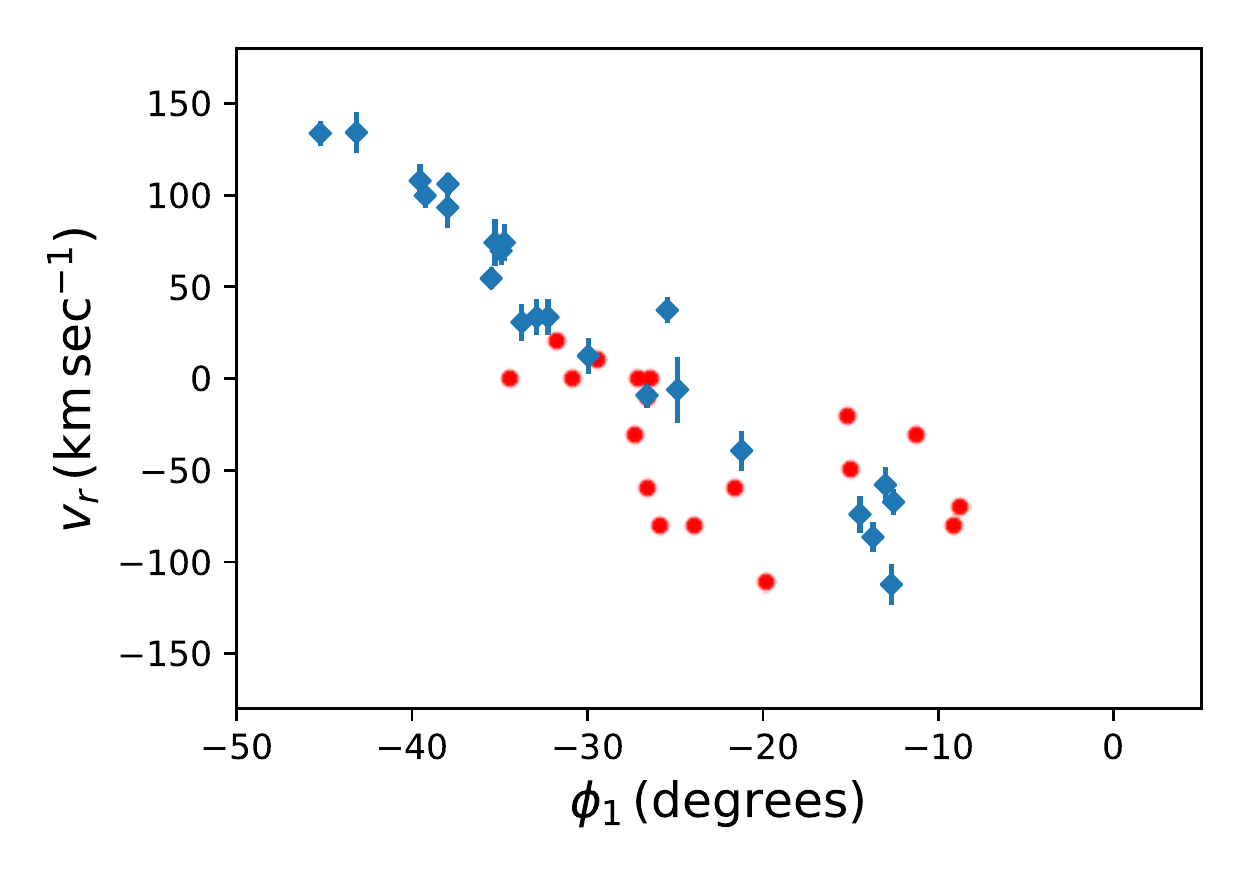}
}
\end{center}
\vspace{-0.5cm}
\caption{Retrieving the missing phase-space information of streams stars with \texttt{STREAMFINDER}. The red dots represent the radial velocity solutions of the GD-1 signal stars that are derived as a by-product of the application of the algorithm. The blue markers are the observed radial velocities of GD-1 stars as tabulated by \citet{Koposov2010}, corrected for the radial component of the Solar reflex motion (taking $V_o = 220\kms$). The \texttt{STREAMFINDER} sampled orbits in radial velocity space at intervals of 10$\kms$ (which effectively causes an uncertainty of 10$\kms$ on the red dots). The agreement with the observations illustrates the power of our algorithm in predicting the missing phase-space information of stream stars.}
\label{fig:comparison}
\end{figure}

\section{Retrieving missing phase-space information from \texttt{STREAMFINDER}}\label{sec:Radial_velocity}

As \texttt{STREAMFINDER} integrates orbits in order to detect stream structures, the primary by-product that the algorithm naturally returns are the possible set of orbital solutions along which the stream might lie. We highlight the power of this by-product by comparing the possible set of radial velocity solutions of the highest likelihood GD-1 stars we obtain from \texttt{STRAMFINDER} with the radial velocities of possible GD-1 members listed in Table~1 of \cite{Koposov2010}. The comparison is shown in Figure~\ref{fig:comparison} that displays the agreement between the predicted and the observed stellar velocity measurements. This analysis shows that our algorithm has potential not only for detecting streams, but also for predicting the missing phase-space information of the stream stars. 

Radial velocity and distance information will be missing for the great majority of halo stars in the Gaia DR2 (and successive Gaia catalogues). However, since our algorithm gives the possible orbital solutions for the detected stream structures, it therefore provides a means to complete the 6D phase-space solutions that are possible for a given stream star.

% %%%%%%%%%%%%%%%%%%%%%%%%%%%%%%%%%%%%%%%%%%%%%
% % Section
% %%%%%%%%%%%%%%%%%%%%%%%%%%%%%%%%%%%%%%%%%%%%%

\section{Discussion and Conclusions}\label{sec:Discussion_and_Conclusions}

In this contribution, we have presented the application of our \texttt{STREAMFINDER} algorithm onto the PS1 proper motion dataset in order to detect the GD-1 stream. We chose to analyse a magnitude-limited sample with ${\rm r_{P1}} < 18.5$ which removes most of the MSTO stars of the GD-1 stream, while still containing the stars with well-measured proper motions. While this trimmed sample leads to what is effectively a non-detection with a matched filter search, the application of \texttt{STREAMFINDER} onto the very same data readily shows up the stream at a significance level of $>4.4\sigma$. This both validates our algorithm and the proper motion measurements in the PS1 catalogue.

In addition, we also confirm the presence of a parallel stream-like structure that appears in the neighbouring region of GD-1 at a significance level comparable to that of GD-1, initially suggested by \citet{GrillmairGD12006}. The similarities in the distance and kinematic properties of GD-1 and the parallel stream are striking, and indeed, they currently appear to be converging towards $\alpha\sim 154\deg$, $\delta\sim 40\deg$. It will be interesting to compare the metallicity of GD-1 and the parallel feature, and to study their orbits in detail with proper motions from Gaia DR2 together with accurate radial velocities. 

In a very recent paper, \cite{WhelanBonacaGD12018} suggested that the progenitor of GD-1 lies at $\phi_1=-15\deg$, based on the over-density of stars that they obtain in that region (see their Figure 1). To some extent, the evidence presented here also advocates a similar position for the GD-1's progenitor (see the kink feature in Figure \ref{fig:Our_MF_Grillmair_like}d and the overdensity of GD-1 stars in Figure \ref{fig:STREAMFINDER} at $\phi_1 \sim -18\deg$). This seems to make the stream-fanning origin for the parallel structure somewhat less plausible as the fanning is expected to cause a spreading of the tidal arms at locations along the stream away from the progenitor \citep{Pearson2015fanning}. However in another recent study, \cite{Boer2018} suggested that the GD-1 progenitor is located at the position of an under-density in their MF map at $\phi_1=-45\deg$ ($\alpha\sim 146\deg$, $\delta\sim 32\deg$, our coordinate conversion) which is surrounded by overdense stream segments on either side. If their interpretation is correct, the region displayed in Figure~\ref{fig:STREAMFINDER} is fully occupied by the trailing stream. Given the similar distances, orbits, and stellar populations of GD-1 and the parallel structure, the spatial configuration shown in Figure~\ref{fig:STREAMFINDER}d is therefore highly suggestive of stream-fanning (\citealt{Pearson2015fanning}, cf. their Figure~4). The fanning-out of the orbits of the stream could be provoked by the triaxiality of the bar; it will be interesting in future work to simulate the dynamical evolution of the GD-1 progenitor given these new observational constraints. However, at the present time we cannot rule out the alternative possibility that the parallel structure is a new stellar stream formed from a different progenitor than that of GD-1. 

The positive detection of GD-1 in this PS1 proper motion sample with the \texttt{STREAMFINDER} suggests that it will be possible to find other similar structures in Gaia DR2, where the proper motion uncertainties of stars will be a factor of $\sim 5$ better in each proper motion dimension (yielding a $\sim 25$ times better-resolved phase-space volume). Later Gaia releases are expected to further improve the astrometric accuracy by more than a factor of 5.

Our algorithm naturally delivers the possible set of orbital solutions of the detected stream structures. Our analysis here shows good agreement between the radial velocities of the GD-1 stars obtained as a by-product from \texttt{STREAMFINDER} and from spectroscopic observations. This missing phase-space information that the algorithm provides may be used to estimate the distribution function of Milky Way streams to some extent, and hence probe the nature and formation history of these star streams and the Galactic halo that together they span.

% %%%%%%%%%%%%%%%%%%%%%%%%%%%%%%%%%%%%%%%%%%%%%
% % Section
% %%%%%%%%%%%%%%%%%%%%%%%%%%%%%%%%%%%%%%%%%%%%%
\section*{Acknowledgements}

The authors would like to thank Dr. C. J. Grillmair, the reviewer of the paper, for  very useful comments that contributed to the clarity and overall improvement of the paper.

The Pan-STARRS1 Surveys (PS1) and the PS1 public science archive have been made possible through contributions by the Institute for Astronomy, the University of Hawaii, the Pan-STARRS Project Office, the Max-Planck Society and its participating institutes, the Max Planck Institute for Astronomy, Heidelberg and the Max Planck Institute for Extraterrestrial Physics, Garching, The Johns Hopkins University, Durham University, the University of Edinburgh, the Queen's University Belfast, the Harvard-Smithsonian Center for Astrophysics, the Las Cumbres Observatory Global Telescope Network Incorporated, the National Central University of Taiwan, the Space Telescope Science Institute, the National Aeronautics and Space Administration under Grant No. NNX08AR22G issued through the Planetary Science Division of the NASA Science Mission Directorate, the National Science Foundation Grant No. AST-1238877, the University of Maryland, Eotvos Lorand University (ELTE), the Los Alamos National Laboratory, and the Gordon and Betty Moore Foundation.

%%%%%%%%%%%%%%%%%%%%%%%%%%%%%%%%%%%%%%%%%%%%%%%%%%



\bibliographystyle{mnras}
\bibliography{ref1} % if your bibtex file is called example.bib

%%%%%%%%%%%%%%%%%%%%%%%%%%%%%%%%%%%%%%%%%%%%%%%%%%

% Don't change these lines
\bsp	% typesetting comment
\label{lastpage}
\end{document}